\documentclass[leqno,10pt,a4paper,dvipdfmx]{article}
\usepackage{aer}
\usepackage{aer-osborn}
\usepackage{graphicx}
\usepackage[bf,hang,margin=7truemm]{caption}[2007/04/09]
\usepackage{amsmath}
\usepackage[]{natbib}
\def\figkokopdf#1#2#3#4{
\begin{figure}[!tp]
\begin{center}
\includegraphics[width=#3\textwidth,bb=#4]{#1.pdf}
\caption{#2}
\label{fig:#1}
\end{center}
\end{figure}}
\begin{document}
\def\addnum#1{$^#1$}
\title{Productivity Dispersion:\\ Facts, Theory, and Implications}
\author{Hideaki Aoyama$^1$\footnote{Corresponding author: 
{\tt hideaki.aoyama@scphys.kyoto-u.ac.jp}.},
Hiroshi Yoshikawa$^2$,
Hiroshi Iyetomi$^3$,\\[5pt]
and Yoshi Fujiwara$^{4}$
}
\maketitle
\vspace{-25pt}
\begin{center}
{\small\sl
\addnum1 
Department of Physics, Kyoto University, Kyoto 606-8501, Japan\\[3pt]
\addnum2 
Faculty of Economics, University of Tokyo, Tokyo 113-0033, Japan\\[3pt]
\addnum3 
Department of Physics, Niigata University, Niigata 950-2181, Japan\\[3pt]
\addnum4 
NiCT/ATR CIS, Applied Network Science Lab., Kyoto 619-0288, Japan}
\end{center}

\begin{abstract}
We study productivity dispersions across workers, firms and 
industrial sectors.
Empirical study of the Japanese data
shows that they all obey the Pareto law, and also that the Pareto index decreases with the level of aggregation.
In order to explain these two stylized facts, we propose a theoretical framework
built upon the basic principle of statistical physics.  
In this framework, we employ the  concept of superstatistics which accommodates fluctuations of aggregate demand.
\end{abstract}

\centerline{JEL No: E10, E30, O40}

\centerline{KUNS-2140}

\def\muf{{\mu_{\rm F}}}
\def\mue{{\mu_{\rm W}}}
\def\mus{{\mu_{\rm S}}}
\def\ps{P^{(\rm S)}(c)}
\def\psc{P^{(\rm S)}_>(c)}
\def\pf{P^{(\rm F)}(c)}
\def\pfc{P^{(\rm F)}_>(c)}
\def\pe{P^{(\rm W)}(c)}
\def\pec{P^{(\rm W)}_>(c)}
\def\pfd{P^{(\rm F)}(0)}
\def\pfe{P^{(\rm F)\prime}(0)}
\def\wn{w_{\{n\}}}
\def\pn{P_{\{n\}}}
\def\sumk{\sum_{k=1}^K}
\def\pk{p_k}
\def\dtil{\widetilde{D}}
\def\dc{\Delta_c}
\def\act{\langle c \rangle_{\beta}}
\def\actn#1{\langle c^#1 \rangle_{\beta}}
\def\ac{\langle c \rangle_{0}}
\def\acn#1{\langle c^#1 \rangle_{0}}
\section{Introduction}

The standard economic analysis postulates that the marginal products of a production factor 
such as labor are all equal across firms, industries, and sectors in equilibrium. 
Otherwise, there remains a profit opportunity, and this contradicts to the notion of equilibrium.  
Factor endowment, together with preferences and technology, determines equilibrium in such a way 
that the marginal products are equal across sectors and firms.  

A bold challenge to this orthodox theory was given by \citet{k1936}.  
He pointed out that the utilization of production factors is not full, and, therefore, 
that factor endowment is not an effective determinant of equilibrium.  
In \textit{the General Theory}, Keynes identified the less than full utilization of production factor 
with the presence of involuntary unemployment of labor.  Much controversies revolved around 
the theoretically ambiguous notion of ``involuntary'' unemployment.  
However, for Keynes' economics to make sense, the existence of involuntary unemployment is 
not necessary. 
More generally, it is enough to assume that there is {\it underemployment} in the economy in the sense 
that the marginal products are not uniform at the highest level.  
In effect, what Keynes said is that in the demand-constrained equilibrium, 
productivity of  production factor differs across industries and firms.  

There are, in fact, several empirical findings which strongly suggest that productivity dispersion 
exists in the economy.  
The celebrated Okun's Law is an example.  
The standard assumption on the neoclassical production function entails that the elasticity of 
output with respect to labor input (or the unemployment rate) is less than one.  
\citet{okun}, however, found this elasticity to be three for the U.S. economy.  
This finding turns out to be so robust that it has eventually become known as the Okun's Law.  

Okun attempts to explain his own finding by resorting to several factors.  
They are 
(a) additional jobs for people who do not actively seek work in a slack labor market 
but nonetheless take jobs when they become available; 
(b) a longer workweek reflecting less part-time and more overtime employment; 
and 
(c) extra productivity.  
On cyclical changes in productivity, he argues as follows: 

\begin{quotation}
``I now believe that an important part of the process involves a downgrading of labor in a slack 
economy---high-quality workers 
avoiding unemployment by accepting low-quality and less productive jobs.  
The focus of this paper is on the upgrading of jobs associated with a high-pressure economy.  
Shifts in the composition of output and employment toward sectors and industries 
of higher productivity boost aggregate productivity as unemployment declines.  
Thus the movement to full employment draws on a reserve army of the underemployed as well as of 
the unemployed.  In the main empirical study of this paper, 
I shall report new evidence concerning the upgrading of workers into more productive jobs 
in a high-pressure economy. (Okun (1962, p.208))''
\end{quotation}
Okun well recognized and indeed,
by way of his celebrated law,
demonstrated that there exists dispersion of labor 
productivity in the economy. 

Another example is wage dispersion, an observation long made by labor economists.  
\citet{mortensen}, for example, summarizes his analysis of wage dispersion as follows.

\begin{quotation}
``Why are similar workers paid differently?  Why do some jobs pay more than others?  
I have argued that wage dispersion of this kind reflects differences in employer 
productivity. \dots 
Of course, the assertion that wage dispersion is the consequence of productivity dispersion 
begs another question.  
What is the explanation for productivity dispersion?  (Mortensen (2003, p.129))''
\end{quotation}

To this question, Mortensen's explanation is as follows:

\begin{quotation}
``Relative demand and productive efficiency of individual firms are continually shocked by events. 
The shocks are the consequence of changes in tastes, changes in regulations, and changes induced 
by globalization among others.  
Another important source of persistent productivity differences across firms 
is the process of adopting technical innovation.  
We know that the diffusion of new and more efficient methods is a slow, drawn-out affair.  
Experimentation is required to implement new methods.  
Many innovations are embodied in equipment and forms of human capital that are necessarily 
long-lived.  
Learning how and where to apply any new innovation takes time and may well be highly firms 
specific. 
Since old technologies are not immediately replaced by the new for all of these reasons, 
productive efficiency varies considerably across firms at any point in time.  
(Mortensen (2003, p.130))''
\end{quotation} 

It is important to recognize that the processes Mortensen describes are intrinsically stochastic.  
Because there are millions of firms in the economy, 
it is absolutely impossible for economists to pursue the precise 
behavior of an individual firm although no doubt each firm tries 
to maximize profits, perhaps dynamically, under certain constraints.  
Therefore, we must explore such stochastic dynamics  in the economy as a whole 
by different method.  In this paper, we pursue such a stochastic approach.  

The ``stochastic approach'' was indeed once popular in diverse areas of study such as 
income distribution and firm (city) size: Primary examples are \citet{cham1973} and \citet{ijiribook}.
However, the stochastic approach eventually lost its momentum.  
According to \citet{sutton}, the reason is as follows:

\begin{quotation}
``It seems to have been widely felt that these models might fit well, but were ``only stochastic.''  
The aim was to move instead to a program of introducing stochastic elements into conventional maximizing models.
(Sutton (1997, p.45))''
\end{quotation}

This trend is  certainly in accordance with  the motto of the mainstream macroeconomics.
However, Sutton, acknowledging the importance of the stochastic approach, 
argues that\footnote{Incidentally, \citet{sutton}, 
evidently having the central limit theorem in mind, 
identifies what he calls ``purely statistical'' effects with the effects of \textit{independent} stochastic variables.  
However, the method of statistical physics is effective even when stochastic variables are correlated, 
that is they are \textit{not} independent.} 
\begin{quotation}
``a proper understanding of the evolution of structure may require an analysis not only of such economic mechanisms, 
but also of the role played by purely statistical (independence) effects, and that a complete theory will need to find an appropriate way of combining these two  strands. (Sutton (1997, p.57))''
\end{quotation}
Actually, 
as the system under investigation becomes larger and more complicated, 
the importance of the stochastic approach increases.  
In fact, it is almost the definition of macro-system to which the basic approach based on statistical physics can be usefully applied.  
In natural sciences such as physics, chemistry, biology and ecology, the ``stochastic approach'' is routinely taken to analyze macro-system.  Why not in macroeconomics?

In what follows, we first explain the concept of \textit{stochastic macro-equilibrium}.
Next, we present an empirical analysis of productivity dispersion across the Japanese firms.
Then, we present a theory which can explain the observed distribution of productivity.  
The theoretical analysis which we offer follows the standard method of statistical physics, 
and abstracts from microeconomic analysis of the behavior of individual agent.  
We believe that as Sutton (1997) suggests, our analysis complements the standard search theoretical approach
which pays careful attention to microeconomic behavior.  
Finally, we discuss implications of our analysis and a possible direction of future research.  

\section{The Stochastic Macro-equilibrium}
\citet{tobin} proposed the concept of \textit{stochastic macro-equilibrium} 
in his attempt to explain the observed Phillips curve.  He argues that 

\begin{quotation}
``[it is] stochastic, because random
intersectoral shocks keep  individual labor markets in diverse states of disequilibrium; 
macro-equilibrium, because the perpetual flux of particular markets produces fairly 
definite aggregate outcomes.  (Tobin (1972, p.9))''
\end{quotation}
His argument remained only verbal.  However, as we will see it shortly, 
the fundamental principle of statistical physics, in fact, provides the exact foundations for the concept 
of macro-equilibrium.  A case in point is productivity dispersion in the economy.  

We consider the economy in which there are $K$ firms with their respective
productivities $c_1, c_2, \dots.$ 
Without loss of generality, we can assume
\begin{equation}
 \qquad c_1<c_2<\dots < c_K.
\end{equation}
Labor endowment is $N$.   
We distribute $n_k$ workers to the $k$-th firm.
Thus, the following equality holds:
\begin{equation}
\sumk n_k =N.
\end{equation}

The neoclassical theory takes it for granted that $n_K$ is $N$
while $n_k$ ($k \neq K)'$s are all zero.
Instead, in what follows, we seek the most probable distribution of productivity across
workers under suitable constraints.\footnote{The productivity $c_k$
corresponds to allowed energy level, and workers to
distinguishable particles in statistical physics.}
The possible number of a particular configuration $(n_1, n_2, \dots, n_K), \wn$ is
\begin{equation}
\wn=\frac{N!}{\prod_{k=1}^K n_k!}.
\label{baai}
\end{equation}
Because the total number of possible configurations is
$K^N$, the probability of occurrence of such a configuration,  $\pn$,
on the assumption of equal probabilities for all configurations, is
given by
\begin{equation}
\pn=\frac1{K^N}\frac{N!}{\prod_{i=1}^K n_i!}.
\label{pn}
\end{equation}
Following the fundamental principle of statistical physics,
we postulate that the configuration $\{n_1, n_2, \dots, n_K\}$
that maximizes $\pn$ under suitable constraints is realized in equilibrium.
The idea is similar to the method of maximum likelihood in statistics or econometrics.

It is extremely important to recognize that this analysis is consistent with the standard 
assumption in economics that economic agents maximize their objective functions. 
As each particle satisfies the Newtonian law of motion, and maximizes the Hamiltonian 
in physics, in the economy, to be sure, all the micro agents optimize.  
However, constraints facing these micro agents, 
and even their objective functions keep changing due to idiosyncratic shocks.  
The situation is exactly the same as in physics where we never know the initial conditions for all the particles.  
In the economy, the number of micro agents is less than the counterpart 
in physics which is typically $10^{23}$, but still there are $10^{6}$ firms and $10^{7}$ households!  
It is simply impossible and meaningless to analyze micro behavior of agent in great detail.  
For studying macro-system, we must take behaviors of micro agents as stochastic even though their behaviors are purposeful.  
It is the fundamental principle of statistical physics 
that we observe the state of macro-system which maximizes the probability, (\ref{pn}).

Toward our goal, we define the following quantity; 
\begin{equation}
S=\frac{\ln \pn}{N}+\ln K
\simeq
\frac1N \left(N\ln N - \sumk n_i\ln n_i\right) 
=-\sumk \pk \ln \pk,
\label{entropy}
\end{equation}
where $\pk$ is defined as
\begin{equation}
\pk\equiv \frac{n_k}{N}.
\end{equation}
Here, we assume that $N$ and $n_k$'s are large, and apply the
Stirling formula:
\begin{equation}
 \ln m! \simeq m\ln m -m \quad \mbox{for } m \gg 1.
\label{mm}
\end{equation}
The quantity $S$ corresponds to the Boltzmann-Gibbs entropy.
Note that the maximization of $S$ is equivalent to that of $\pn$.

We maximize $S$ under two constraints.
One is the normalization condition
\begin{equation}
\sumk p_k =1.
\label{norm}
\end{equation}
This is, of course, equivalent to the resource constraint, 
$\sumk n_k=N.$
The other constraint requires that the total output (GDP) $Y$ 
is equal to the aggregate demand $\dtil$,
\begin{equation}
N \sumk c_k p_k = Y = \dtil.
\label{dn}
\end{equation}
For convenience, we define aggregate demand relative to factor endowment,
$D$ as follows:
\begin{equation}
D=\frac{\dtil}{N}.
\end{equation}
In what follows, we simply call $D$ the aggregated demand.
As we will shortly see it, 
the aggregate demand, $D$ determines the state of stochastic macro-equilibrium.

We maximize the following Lagrangian form:
\begin{equation}
S-\alpha\left(\sumk \pk-1\right)
-\beta \left(\sumk c_k p_k- D\right).
\label{exme}
\end{equation}
Differentiating (\ref{exme}) with respect to $p_i$,
we obtain
\begin{equation}
\ln \pk +(1+\alpha) + \beta c_k=0.
\end{equation}
This yields
\begin{equation}
\pk=e^{-(1+\alpha)} \, e^{-\beta c_k}.
\end{equation}
The normalization condition or the resource constraint, (\ref{norm}) determines $\alpha$ so that
the distribution we seek is obtained as follows:
\begin{equation}
\pk=\frac1{Z(\beta)} \, e^{-\beta c_k}.
\label{boltz}
\end{equation}
Thus, the distribution which maximizes the probability, (\ref{pn})
under two constraints, (\ref{norm}) and (\ref{dn}) is exponential. 
We may call it \textit{the equilibrium distribution}.
\textit{The exponent of the equilibrium distribution is the Lagrangian multiplier $\beta$ corresponding to 
the aggregate demand constraint in (\ref{exme})}. 

This exponential distribution 
is called
\textit{the Maxwell-Boltzmann distribution}.
Here, $Z(\beta)$ is what is called the partition function in physics,
and makes sure that $p_k$'s sum up to be one. 
\begin{equation}
Z(\beta)\equiv \sumk e^{-\beta c_k}.
\label{parti}
\end{equation}

The exponent $\beta$ is equivalent to 
the inverse of \textit{temperature}, $1/T$
in physics.
Equations (\ref{dn}) and (\ref{boltz}) yield the following:
\begin{equation}
D=
\frac1{Z(\beta)} \, \sumk c_k e^{-\beta c}
=-\frac{d}{d\beta}\ln Z(\beta).
\label{dnbeta}
\end{equation}
This equation relates 
the aggregate demand, $D$ to the exponent of the distribution $\beta$
by way of the partition function, $Z(\beta)$.
Note that at this stage, the distribution of productivity $c_k$
is arbitrary. Once it is given, the partition function
$Z(\beta)$ is defined by equation(\ref{parti}),
and it, in turn, determines\footnote{This is exactly the same as 
in the standard analysis in statistical physics where the relation between 
the average energy ($D$) and the temperature ($T=1/\beta$) is 
determined once the energy levels ($c_k$) are given for the system under 
investigation.} the relation between $D$ and $\beta$.

\figkokopdf{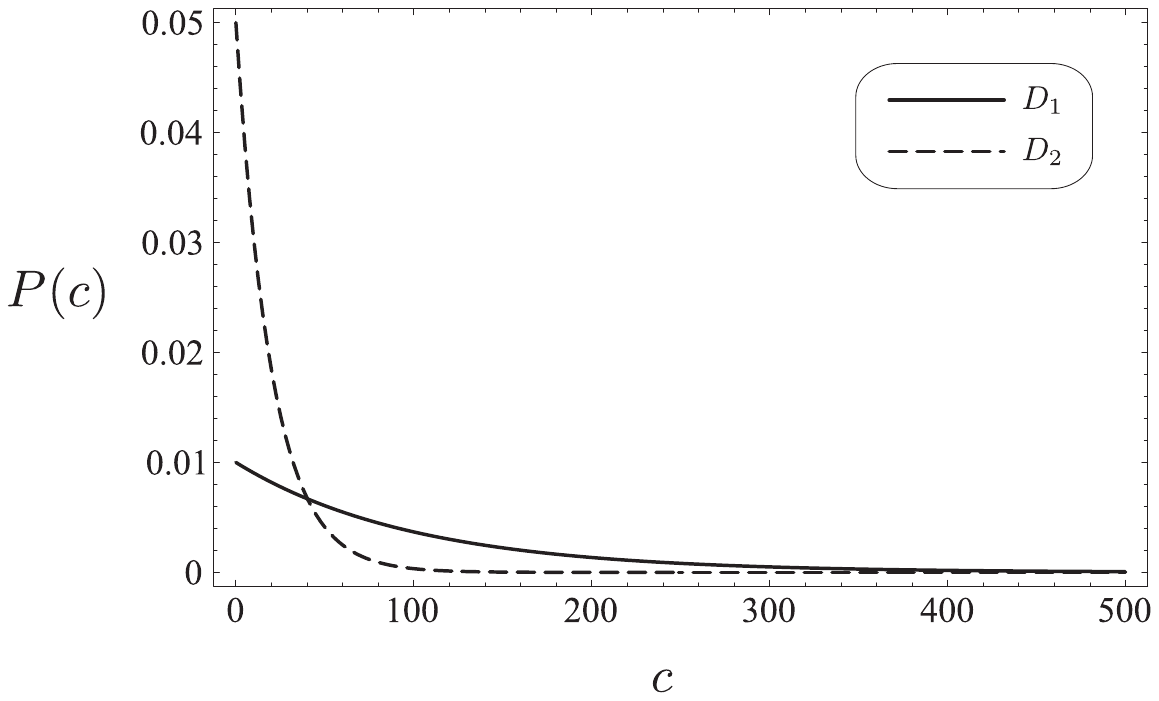}{\textsc{Productivity Distribution Predicted by 
the Stochastic Macro-equilibrium Theory 
for Two Different Values of the Aggregate Demand, $D$, $D_1>D_2$}
{\it Notes:} The values of $\beta$ are obtained from 
equation (\ref{ayrel}) for $D_1=20$ and $D_2=100$.}{0.7}{256 200 585 397}

Suppose that the distribution of productivity across firms 
is {\sl uniform}, that is, $c_k=k \dc$, where $\dc$ is a constant.
Then, from (\ref{boltz}), we obtain 
\begin{equation}
Z(\beta)=\sum_{k=1}^K e^{-\beta k \dc}
=e^{-\beta \dc} \frac{1-e^{-\beta(K-1)\dc}}{1-e^{-\beta \dc}}
\simeq \frac{1}{\beta \dc}, 
\end{equation}
under the assumption that $\beta \dc \ll 1$ and $\beta K \dc \gg 1$.
Therefore, in this case, we know from equations (\ref{parti}) and (\ref{dnbeta}), 
that the exponent of the equilibrium exponential distribution, 
$\beta$ is equal to the inverse of aggregate demand: 
\begin{equation}
D= -\frac{d}{d \beta}\ln Z(\beta) 
= -\frac{d}{d \beta} \ln \left( \frac{1}{\beta \dc} \right)
=\frac{1}{\beta}.
\label{ayrel}
\end{equation}
Because the inverse of $\beta$ is temperature, in this case, 
aggregate demand is equivalent to temperature.  

In summary, under the assumption that the productivity dispersion \textit{across firms} is uniform, 
the equilibrium distribution of productivity \textit{across workers} becomes exponential, 
namely the Maxwell-Boltzmann distribution with the exponent equal to the inverse of the 
aggregate demand, $D$ (Yoshikawa (2002), and Aoki and Yoshikawa (2007)):
\begin{equation}
p_k  = \frac{n_k}{N} = \frac{e ^{-{\frac{c_k}{D}}}}{\Sigma ^{K} _{k=1} e ^{-{\frac{c_k}{D}}}}.
\end{equation}
When the aggregate demand, $D$ is high, 
the distribution becomes flatter meaning that production factors are
mobilized to firms or sectors with high productivity, and \textit{vice versa}.
Figure \ref{fig:ay0} shows  two distributions of productivity corresponding to high aggregate 
demand $D_1$ and low $D_2$.

The present analysis provides a solid foundation for Okun's 
argument that in general, production factors are under-utilized,
 and that workers upgrade into more productive jobs in a ``high-pressure'' economy.
So far, so fine.
However, the Boltzmann-Gibbs distribution is precisely exponential.
Does it really fit our empirical observation?
We next turn to this question.

\section{Stylized Facts}
\def\cm{c_{\rm M}}
\def\cmz{c_{{\rm M}0}}
\def\pa{P_\alpha(\alpha)}
\def\pm{P_{\rm M}(\cm)}
\def\mucm{\mu_{\rm M}}
\def\pca{P_{c,\alpha}(c,\alpha)}
\def\pc{P(c)}
\def\pcon{P(\alpha\,|\,c)}
In this section, we empirically explore how productivity
is actually distributed.  
Before we proceed,
we must point out
that what we observe is the average productivity $c$,
defined by $c=Y/L.$
where $Y$ is the output, and $L$, the labour input.
What matters theoretically is, of course,
the unobsorbed \textit{marginal} productivity $\cm$ defined by
\begin{equation}
\cm = \frac{\partial Y}{\partial L}.
\end{equation}
We will shortly discuss the relationship between the distributions of these two
different productivities.
There, we will see that the 
difference between $c$ and $\cm$ does not affect our results on the
empirical distributions.

Another problem is that we measure $L$ in terms of the number of workers, as such 
data is readily available. One might argue that theoretically, it is more desirable
to measure $L$ in terms of work-hours, or for that matter, even in terms of work
efficiency unit.
The effects of this possible ``measurement errors'' can also be handled 
in a similar way which we will explain in subsection B.
That is, we can safely ignore the measurement error problem as well for our present purpose.

\subsection{Empirical Distributions}
We use the Nikkei-NEEDS database, which 
is a major representative database for the listed firms in Japan.
The details of this database is given in \ref{appdata}, together with
a brief description of the data-fitting.

We study productivity distributions at three aggregate levels, namely, across
workers, firms, and industrial sectors.
In order to explain empirical distributions, we find continuous
model more convenient than discrete model. 
Accordingly, we define, for example, the probability density function of firms with productivity 
$c$ as $\pf$.
The number of firms with productivity between $c$ and $c+dc$ is
now $K\pf dc$. 
It satisfies the following normalization condition:
\begin{equation}
\int_0^\infty \pf \,dc=1.
\label{pfnorm}
\end{equation}

In section II, 
we explained that equilibrium productivity dispersion \textit{across workers} becomes the
exponential distribution under the assumption that
the distribution
of productivity \textit{across firms} is uniform.
Here, $\pf$ is not restricted to be uniform.
Rather the distribution $\pf$ is to be determined empirically.
We analogously define $\pe$ for workers, and $\ps$ for industrial sectors.
We denote their respective cumulative probability distributions by $P_>^{\rm (S,F,W)}(c)$:
\begin{equation}
P_>^{(*)}(c)=\int_c^\infty P^{(*)}(c')\,dc'\quad(*={\rm S, F, W}).
\label{cumpdef}
\end{equation}
Because the cumulative probability is the probability that a  firm's
(or worker's or sector's)
productivity is larger than $c$, it can be measured by
rank-size plots, whose vertical axis is the [rank]/[the total number of firms]
and the horizontal axis $c$. The rank-size plots has advantages that
it is free from binning problems which haunt probability density
function (pdf) plots and also that it has less statistical noises. 
For these reasons, in what follows, we will show the cumulative probability 
rather than probability density function.

\figkokopdf{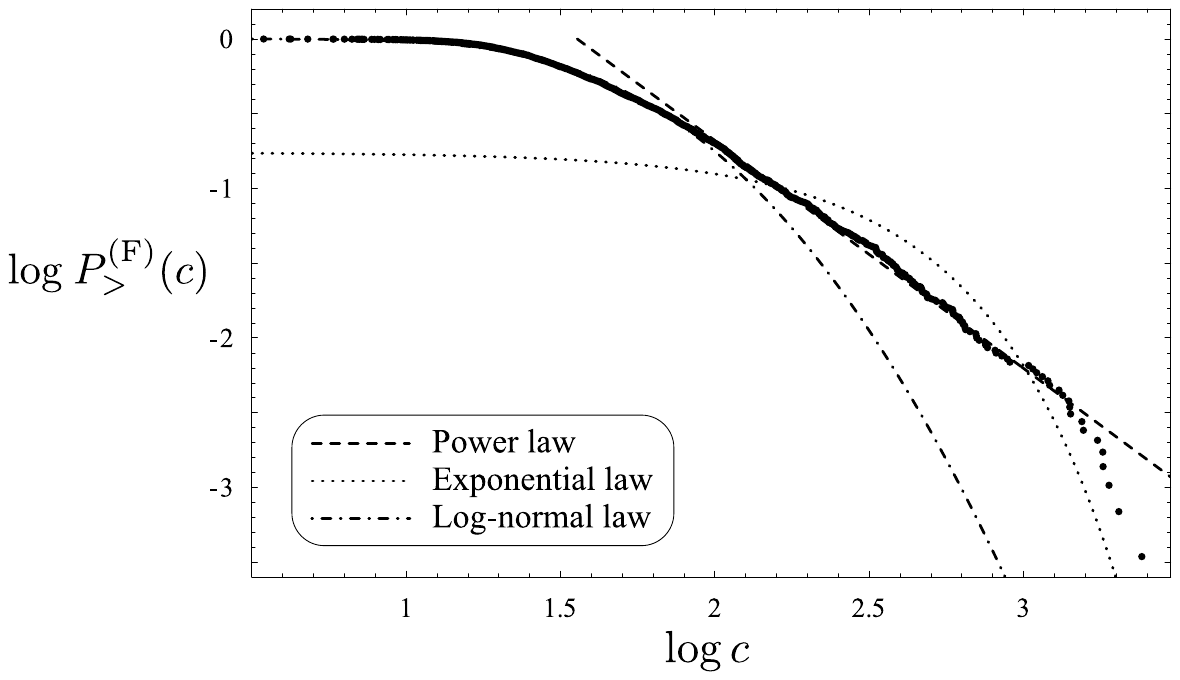}{\textsc{Productivity Distribution across Firms (2005)} 
{\it Notes:} The productivity $c$ is in the unit of $10^6$yen/person.
The best fits for the exponential law and the power law
is obtained for $10<c<3000$.}{0.8}{220 192 617 393}

The productivity distribution across firms, to be exact, 
$\log \pfc$ obtained from the Nikkei-NEEDS data
is plotted in Figure \ref{fig:firm2005_Nikkei} for the year 2005.
The dots are the data points; 
Each dot corresponds to a
firm whose position is determined from its rank and the productivity $c$.
For reference, the power, exponential, and log-normal distributions are shown in the diagram. 
The exponential and log-normal distributions are represented by respective curves while the power-law by a straight line whose slope is
equal to the power exponent.

Evidently, the uniform distribution implicitly assumed in the analysis
in section II does not fit the actual data at all.
For small $c$ (low-productivity), 
the log-normal law (dash-dotted curve) fits well, and 
for large $c$ (high-productivity),  the power-law (broken line) fits well,
with smooth transition from the former to the latter at around $\log c\simeq 2$.
The result  shown in Figure \ref{fig:firm2005_Nikkei} is for the year 2005, but the basically same result holds 
good for other sample periods.  

Cumulative probability $\pfc$ for large $c$ can be, therefore, represented by the following
\textit{power distribution}:
\begin{equation}
\pfc \simeq \left(\frac{c}{c_0}\right)^{-\muf} \quad (c\gg c_0).
\label{firmparetocum}
\end{equation}
where $c_0$ is a parameter that defines the order of 
the productivity $c$.\footnote{We note that $c_0$ has the
same dimension as $c$, so that $\pfc$ is dimensionless, as it should be.}
The power exponent $\muf$ is called \textit{the Pareto index}, and $c_0$, the Pareto scale.
The probability density function is then given by the following:
\begin{equation}
\pf =-\frac{d}{dc} \pfc \simeq \muf \frac{c^{-\muf-1}}{c_0^{-\muf}} \quad  (c\gg c_0).
\label{firmpareto}
\end{equation}

\figkokopdf{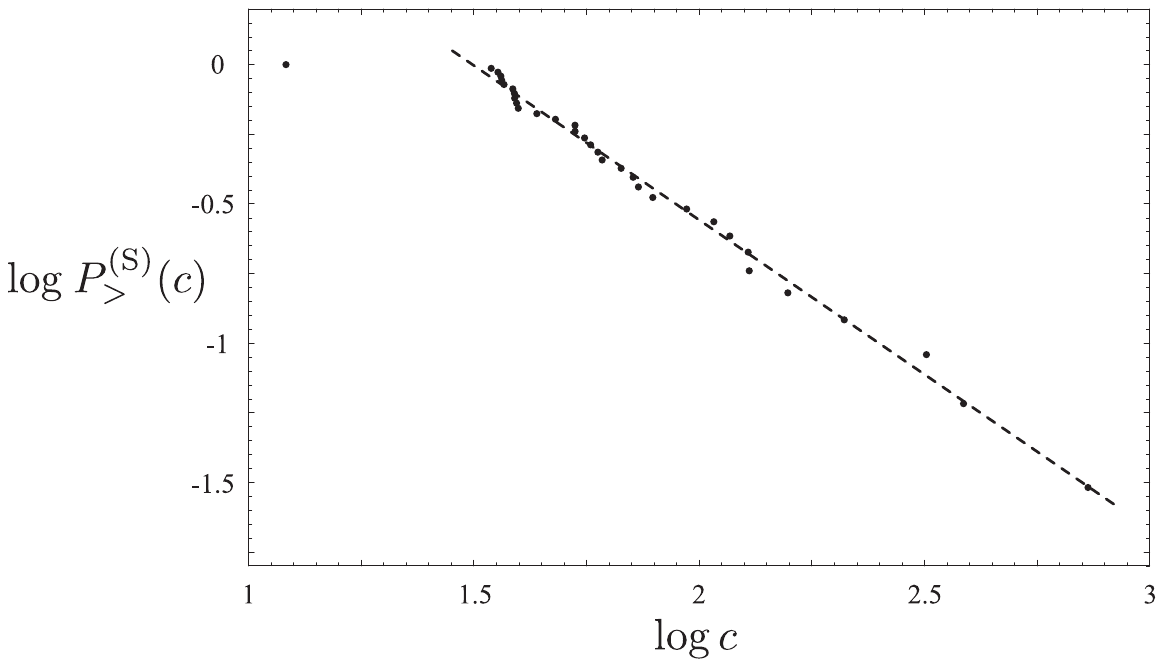}{\textsc{Productivity Distribution across 
{\it Business Sectors} (2005)}}{0.8}{219 198 616 399}
Next, we explore the productivity dispersion \textit{across industrial sectors}. 
The Nikkei-NEEDS database defines 33 sectors.
The productivity distribution across these sectors for the year 2005 is plotted 
in Fig.\ref{fig:sector2005_Nikkei}.
We observe that once again, the power-law (a straight line in the diagram) fits the data pretty 
well.
The same result holds true for all the sample years.

\figkokopdf{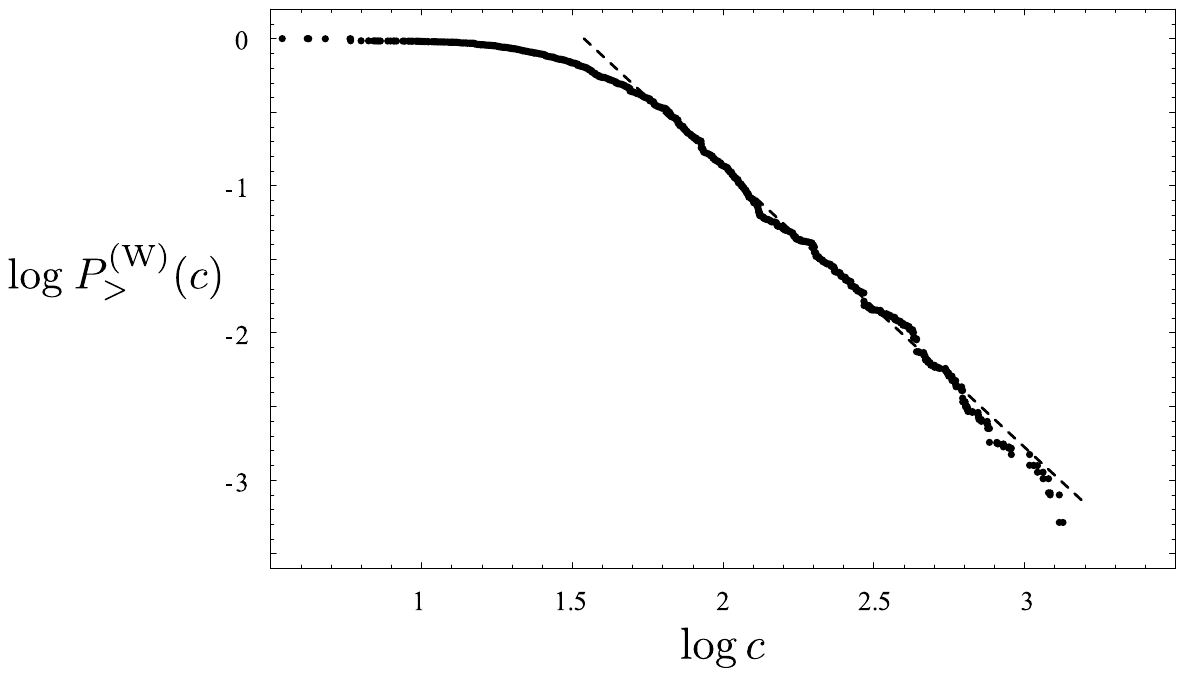}{\textsc{Productivity Distribution across {\it Workers} (2005)}}
{0.8}{422 190 819 391}
Finally, the productivity distribution \textit{across workers}
is plotted in 
Figure \ref{fig:employee2005_Nikkei} for the year 2005.
Again, we observe that the power-law (broken line) fits the data very well for 
large $c$.  A casual observation of 
Figure \ref{fig:employee2005_Nikkei} may make one
wonder whether the share of workers for which
the power-law fits well may be small. This impression is wrong.
In fact, the fitted range is approximately, say, $\log \pec \in [-0.4, -3.1]$,
which translates to the rank of the workers
\begin{align}
[10^{-0.4}, 10^{-3.1}]\times &\mbox{[Total number of workers]}
\simeq [1.52\times10^6,3.04\times10^3]\nonumber
\end{align}
This means that some 1.52 million workers, that is, 39\% of all workers fit the power-law.

The values of the Pareto indices obtained are shown 
in Figure \ref{fig:ParetoIndex} for three levels of aggregation.
We note that the Pareto index decreases as the 
aggregation level goes up from workers to firms, and from firms to the 
industrial sectors.
\figkokopdf{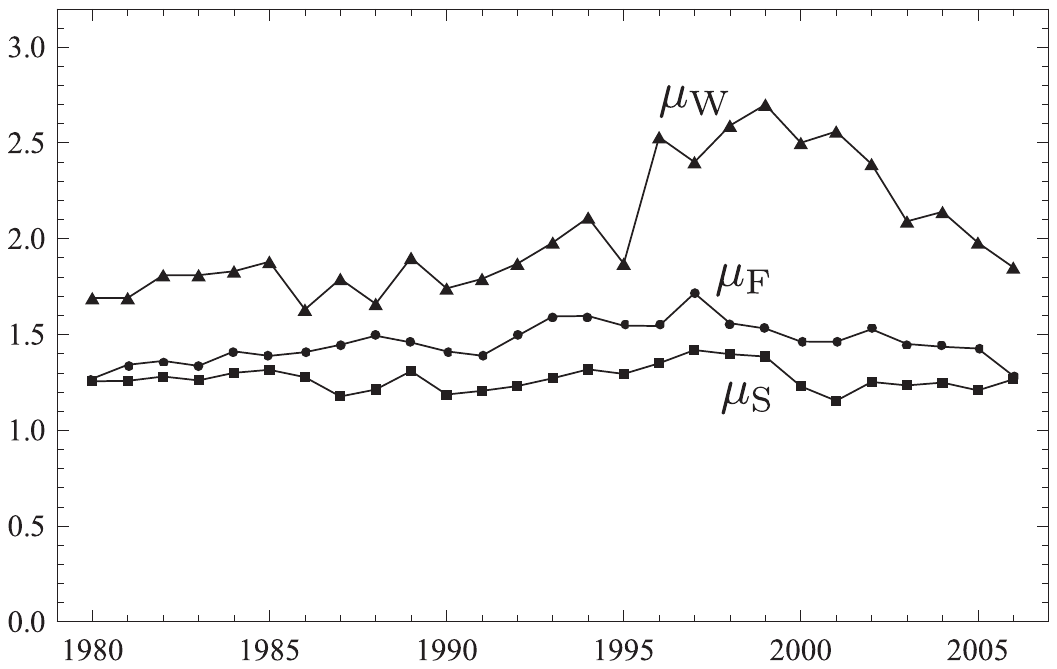}{\textsc{Pareto Indices of the Productivity Distributions 
across Workers, Firms, and Industrial Sectors}}{0.65}{147 325 447 515}

\subsection{Marginal vs.\ Average Productivity}

Before we conclude this section, we digress into
the relationship between the marginal and average productivities of labor.
Theoretically, the marginal productivity matters while what we observe and, 
therefore, used in the above analysis is the average productivity.
To explore the relation between 
the two productivities, $c$ and $\cm$,  
we assume the Cobb-Douglas production function:
\begin{equation}
Y=A K^{1-\alpha} L^\alpha \quad(0<\alpha <1).
\label{cd}
\end{equation}
Equation (\ref{cd}) leads us to the following relation:
\begin{equation}
\cm=\alpha\, c.
\label{cmc}
\end{equation}
In general, the value of $\alpha$ differs across firms.
Therefore, the distribution of $\cm$ is, in general,
different from that of $c$.

However, thanks to ~(\ref {cmc}), 
we can relate the pdf of $\cm$, $\pm$
to the joint pdf of $c$ and $\alpha $, 
$\pca$ as follows:
\begin{align}
\pm &= \int_0^1d\alpha \int_0^\infty dc \, \delta(\cm - \alpha c) \pca
\nonumber\\
&= \int_0^1\frac{d\alpha}{\alpha} \,
P_{c, \alpha}\left(\frac{\cm}{\alpha}, \alpha\right).
\label{myint}
\end{align}
In general, $\pca$ can be written as follows, 
\begin{equation}
\pca=\pf\pcon.
\label{pcon}
\end{equation}
Here, $\pcon$ is the conditional pdf;
it is the pdf of $\alpha$ for the fixed value of productivity $c$.
It is normalized as follows:
\begin{equation}
\int_0^1 \pcon\, d\alpha=1,
\end{equation}
for any value of $c$.
We have already seen that
$\pf$ obeys the power-law
for large $c$:
\begin{equation}
\pf=\muf\frac{c^{-\muf-1}}{c_0^{-\muf}}.
\label{ra}
\end{equation}
From ~(\ref{ra}), ~(\ref{pcon}) and 
~(\ref{myint}), we obtain the following:
\begin{equation}
\pm=\muf\frac{\cm^{-\muf-1}}{c_0^{-\muf}}
\int_0^1d\alpha \,\alpha^{-\mucm}P\left(\alpha\,\left|\,\frac{\cm}{\alpha}\right.\right).
\label{goodres}
\end{equation}
Here, we assumed that $\pcon$ does not extend too near 
$\alpha=0$ so that $\cm/\alpha$ stays in the asymptotic region.
From ~(\ref{goodres}), we can conclude that
if $\alpha$ and $c$ are independent, namely
\begin{equation}
\pcon=\pa,
\end{equation}
then, the distribution of marginal productivity,  $\pm$ also obeys the power law with the \textit{identical}
Pareto index $\muf$ as
for the average productivity:
\begin{equation}
\pm\propto \cm^{-\muf-1}.
\label{samemu}
\end{equation}
In conclusion, 
\textit{to the extent that $\alpha$ and $c$ are independent, the distribution of the unobserved marginal productivity
obeys the power-law with the same Pareto index as
for the observed average productivity}.
A sample Monte-Carlo simulation result is shown in Figure \ref{fig:allp}.
Because this is the log-log plots, the gradient of the straight region
is equal to the Pareto index. 
The equality of the Pareto indices is clearly seen.

In our analysis, $L$ is the number of workers.  Theoretically, it would be
desirable to measure $L$ in terms of work-hours or for that matter,
even in terms of work efficiency.
We can apply the above analysis to the case where the problem is measurement 
error by simply interpreting $\alpha$ in ~(\ref{cmc}) as such measurement error 
rather than a parameter of the Cobb-Douglas production function.
Thus, to the extent 
that the average productivity $c$ and measurement error $\alpha$ are independent,
 the distribution of ``true'' productivity obeys the power-law with the same exponent as for the measured productivity.   
We conclude that the power-laws for productivity dispersion across workers, 
firms and industry obtained above are quite robust.

\figkokopdf{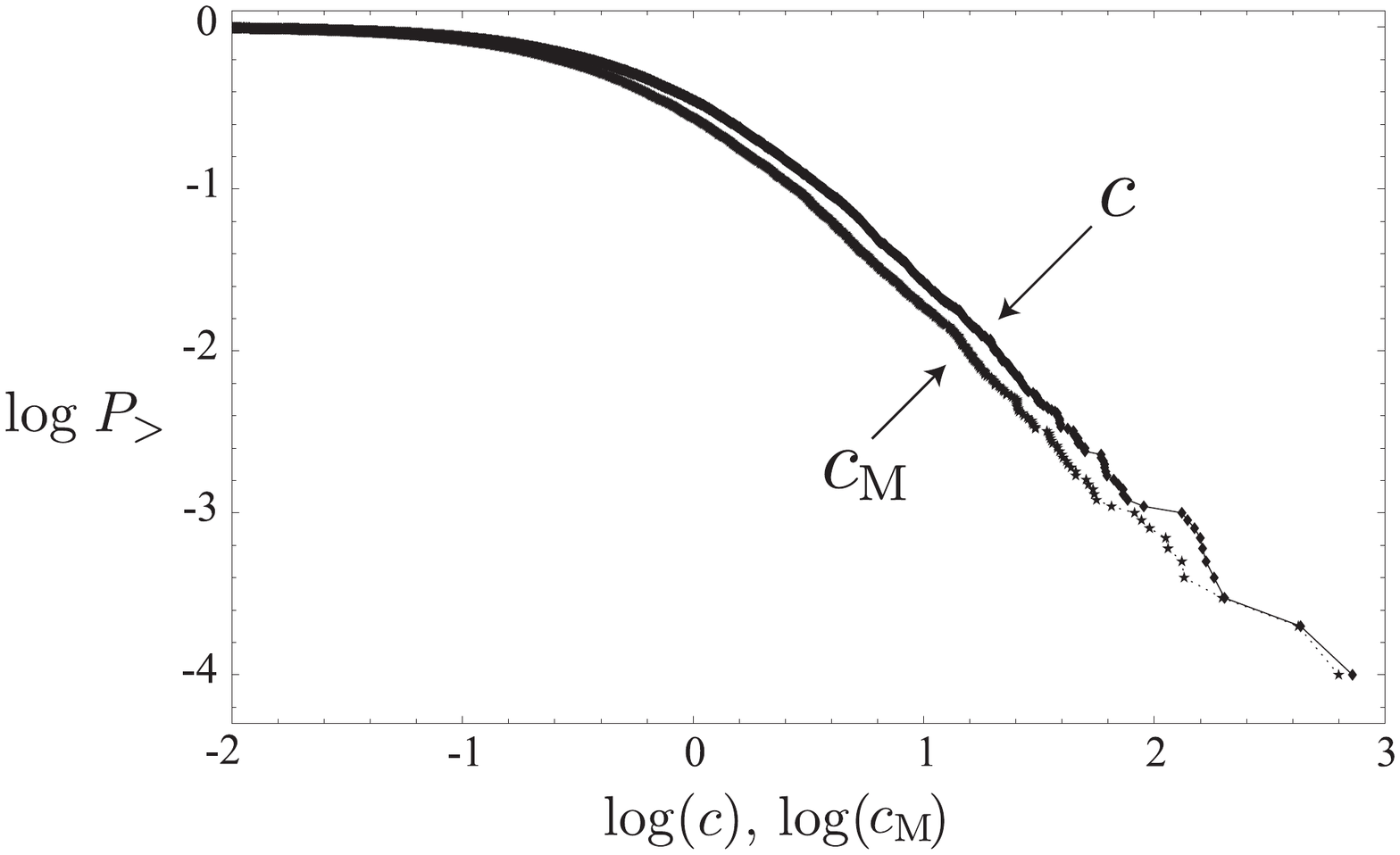}{\textsc{Rank-Size Log-log Plots of the Simulation Results}
{\it Notes:} Distribution for $c$ is for 
the Monte-Carlo data generated from a $\pc$ with $\mucm=1.5$. 
Data for $\cm$ is then created with $\pa$ distributed from $\alpha=0.5$ to 1 uniformly.}{0.6}
{95 106 668 455}

\subsection{Summary of Empirical Observations}
We summarize the empirical observations as two stylized facts:
\begin{enumerate}
\item[I.]
The distribution of productivity obeys the Pareto distribution
(\textit{i.e.} the power-law for the high productivity group)
at every level of aggregation, that is, across workers, firms, and industrial sectors.
\item[II.]
The Pareto index, namely the power exponent decreases as the level of aggregation goes up:
 $\mue > \muf > \mus$ (Figure \ref{fig:ParetoIndex}).\end{enumerate}

As we explained in section II, under the assumption that 
the distribution of productivity \textit{across firms} is uniform leads to the exponential distribution 
of productivity \textit{across workers}.  
Obviously, this model does not fit the empirical observations. 
In the next section, extending the basic framework, we present a theoretical model 
which explains the above stylized facts.

\section{Theory}
In this section, we develop a new theoretical model which explains
two stylized facts in three steps:
First, we discuss the generic framework 
that explains the power-law distribution for firm's productivity. 
Secondly, we extend the basic model explained in section II
by incorporating the stylized  fact that productivity distribution across firms
is not uniform, but rather obeys the power-law.
Because the extended model still fails to explain the stylized facts,
we take the third step;  
We propose the \textit{superstatistics} framework.
It can explain two stylized facts presented in section III.

\subsection{Firm's Productivity Dispersion: A Model of Jump Markov Process}

The standard economic analysis takes it for granted that all the production 
factors enjoy the highest marginal productivity in equilibrium.  
However, this is a wrong characterization of the economy.  The fact is that
production factors cannot be reallocated instantaneously
in such a way that their marginal products are equal in all economic
activities.  Rather, at each moment
in time, there exists a dispersion or distribution of productivity as
shown in the preceding section.
Evidently, an important reason why the marginal products of workers are not equated is, as Mortensen (2003) suggests, that there are differences in productivity across firms.   
To describe the
dynamics of firm's productivity, we employ a continuous-time jump Markov process.
Using the Markov model, we can 
show that the power-law distribution is a generic consequence 
under a reasonable assumption.

Suppose that a firm has a productivity denoted by $c$. In a small
time interval $dt$, the firm's productivity, $c$ increases by a small amount,
which we can assume is unity without loss of generality, with 
probability $w_+(c)\,dt$. 
We denote it as $w_+(c)$ because this probability $w_+$ depends on the level of $c$.
Likewise, it
decreases by unity with 
probability $w_-(c)\,dt$. 
Thus, $w_+(c)$ and $w_-(c)$ are transition
rates for the processes, $c\rightarrow c+1$ and $c\rightarrow c-1$,
respectively. 

We also assume that a new firm is born with a unit of productivity
with probability $p\,dt$. 
On the other hand, a firm
with $c=1$ will be dead if its productivity falls to zero; Thus the
 probability of exit is $w_-(c=1)\,dt$.
A set of the transition rates and the entry probability
specifies the jump Markov process.

Given this Markov model, the evolution of 
the average number of firms
having productivity $c$ at time $t$, $n(c,t)$ obeys the following master equation:
\begin{align}
  \frac{\partial n(c, t)}{\partial t}=&
  w_+(c-1)\,n(c-1,t)+w_-(c+1)\,n(c+1,t) \nonumber \\
 & -w_+(c)\,n(c,t)-w_-(c)\,n(c,t)+p\,\delta_{c,1} \ ,
\label{master}
\end{align}
Here, $\delta_{c,1}$ is 1 if $c=1$ and 0 otherwise.
This equation shows that the change in $n(c,t)$ over time is nothing but the
\textit{net} inflows to the state $c$.

The total number of firms is given by 
\begin{equation}
K_t\equiv\sum_{c=1}^\infty n(c,t).
\end{equation}
We define the aggregate productivity index $C$ as follows:
\begin{equation}
C_t\equiv\sum_{c=1}^\infty c\,n(c,t).
\end{equation}
It follows from (\ref{master}) that
\begin{align}
\frac{d}{dt}K_t &= p-w_-(1)\,n(1,t)
  \label{master_dN}, \\
\frac{d}{dt}C_t &= p-\sum_{c=1}^\infty
  (w_-(c)-w_+(c))\,n(c,t).
  \label{master_dC}
\end{align}

We consider the steady state.  It is the stationary solution of (\ref{master}) such
that $\partial n(c,t)/\partial t=0$. The solution $n(c)$ can be readily obtained by the
standard method.
Setting (\ref{master_dN}) equal to zero, we obtain a
boundary condition that $w_-(1)\,n(1)=p$.
Using this boundary condition, we can easily show
\begin{equation}
  n(c)=n(1)\prod_{k=1}^{c-1}\frac{w_+(c-k)}{w_-(c-k+1)} \ .
\label{statsol}
\end{equation}

Next, we make an important assumption on the transition rates, $w_+$ and $w_-$.
Namely, we assume that the probabilities of 
an increase and a decrease of productivity  depend
on the firms current level of productivity.
Specifically, the higher the current level of productivity is,
the larger a chance of a unit productivity change is.
This assumption means that the
transition rates can be written as
$w_+(c)=a_+\,c^\alpha$ and $w_-(c)=a_-\,c^\alpha$, 
respectively.  Here, $a_+$ and $a_-$ are positive constants, and 
$\alpha$ is greater than 1. 
Under this assumption, the stationary
solution (\ref{statsol}) becomes
\begin{equation}
  n(c)=\frac{n(1)}{1-n(1)/C_{(\alpha)}}
  \frac{(1-n(1)/C_{(\alpha)})^c}{c^\alpha}
  \simeq c^{-\alpha}\,e^{-c/c^*} \ .
\label{statsol1}
\end{equation}
where 
\begin{equation*}
c^*\equiv(n(1)/C_{(\alpha)})^{-1} 
\quad\quad \mbox{and} \quad\quad
C_{(\alpha)}\equiv
\sum_{c=1}^\infty c^\alpha n(c). 
\end{equation*}

We have used the relation
$a_+/a_-=1-n(1)/C_{(\alpha)}$, which follows from 
(37) and (38).  
The approximation in 
(\ref{statsol1}) follows from 
$n(1)/C_{(\alpha)}\ll 1$.
The exponential cut-off works
as $c$ approaches to $c^*$.
However, the value of $c^*$ is practically
quite large. Therefore, we observe the power-law distribution $n(c)\propto
c^{-\alpha}$ for a wide range of $c$ in spite of the cut-off. 
We note that the power
exponent $\mu$ for the empirical distribution presented in section III is related to
$\alpha$ simply by 
\begin{equation*}
\mu=\alpha-1.
\end{equation*}

The present model can be understood easily with the help of an analogy
of the formation of cities. Imagine that $n(c,t)$ is the number of cities
with population $c$ at time $t$. $w_+(c)$ corresponds to a 
birth in a city with population $c$, or an inflow into the city 
from another city. Similarly, $w_-(c)$
represents a death or an exit of a person moving to another city. 
The rates are the instantaneous probabilities that population of
city with the current population $c$ either increases or decreases by
one. They are, therefore, the entry and exit rates of {\it one person\/} times
population $c$, respectively. And a drifter forms his own one-person city with the
instantaneous probability $p$. In this model, dynamics of
$n(c,t)$, namely the average number of cities with population $c$ is given by
equation (34).  
In the case of population dynamics, one
might assume that the entry (or birth) and exit (or death) rates of a person, 
$a_+$ and $a_-$ are independent
of the size of population of the city in which the person lives. Then,
$w_+(c)$ and $w_-(c)$ become linear functions of $c$, namely, $a_+c$ and
$a_-c$. Even in population dynamics, though, one might assume that the
entry rate of a person into big city is higher than its
counterpart in small city because of the better job opportunity or the
attractiveness of  ``city life.''  
The same may hold for the exit and death rates because of congestion or
epidemics.

It turned out that in dynamics of firm productivity, both the ``entry'' and
``exit'' rates of an existing ``productivity unit'' are increasing
functions of $c$, namely the level of productivity a part of which
that particular unit happens to be;  
To be concrete, they are $a_+c$ and $a_-c$. Thus,
$w_+(c)$, for example, becomes $a_+c$ times $c$ which is equal to $a_+c^2$.
Likewise, we obtain $w_-(x)=a_-x^2$.
This is the case of $\alpha=2$, the so-called Zipf law (see Sutton (1997)). 
\footnote{
See \citet{ijiri1975sda,marsili1998} for the formation of cities
whose sizes obey the Zipf law. 
Our present model of productivity dynamics has, in fact, a very close analogy 
to dynamics of city size.  
It is indeed mathematically
equivalent to the model of \citet{marsili1998} which analyzed dynamics of city size, 
except for an additional assumption that $n(1)/C_{(\alpha)}=p$.}

There is also a technical reason why we may expect $\alpha$ to be larger than two.
From (\ref{statsol1}), we can write the total number of firms 
and the aggregate productivity index as
\begin{eqnarray}
  K&=&\frac{n(1)}{\Gamma(\alpha)}
  \int_0^\infty\frac{t^{\alpha-1}}{e^t-1+n(1)/C_{(\alpha)}}dt \ ,
  \label{statN} \\
  C&=&\frac{n(1)}{\Gamma(\alpha-1)}
  \int_0^\infty\frac{t^{\alpha-2}}{e^t-1+n(1)/C_{(\alpha)}}dt \ ,
  \label{statC}
\end{eqnarray}
where $\Gamma(z)$ is the gamma function defined by
$\Gamma(z)=\int_0^\infty t^{z-1}e^t dt$. 
In the limit as $n(1)/C_{(\alpha)}$ goes to zero, 
the integral in (\ref{statN}) is finite for $\alpha>1$,
while the integral in (\ref{statC}) tends to be arbitrarily large
for $1<\alpha<2$.
Therefore, for the finiteness of both (\ref{statN}) and (\ref{statC}), 
it is reasonable to assume that $\alpha\geq 2$.

In summary, under the reasonable assumption 
that the probability of a unit change in productivity is an
increasing function of its current level, $c$, we
obtain power-law distribution 
as we actually observe.
Now,  
economists are prone to take changes in productivity as ``technical progress.''  
That is why the focus of attention is so often on R \& D investment.  
However, if productivity growth is always technical progress, 
its decrease must be ``technical regression,'' the very existence of which one might  question. 
At the firm level, an important source of productivity change is actually a sectoral shift of demand; 
When demand for product A increases, for example, productivity at the firm producing 
A increases,  and \textit{vice versa}.  
\citet{fay} indeed document such changes in firm's labor productivity by way of changes 
in the rate of labor hoarding.  
Stochastic productivity changes which our Markov model describes certainly 
include technical progress, particularly in the case of an increase,
but at the same time, represent \textit{allocative demand disturbances} 
the importance of 
which \citet{davis} have so
persuasively demonstrated in their book entitled \textit{Job Creation and Destruction}.
The empirical observation on productivity dispersion 
and our present analysis suggest that the probability of a unit allocative demand disturbance 
depends on the current size of firm or sector.

\subsection{The Stochastic Macro-equilibrium Once Again}
The basic framework explained in section II presumes that productivity dispersion across firms, namely $\pfc$ is uniform.  However, $\pfc$ actually obeys the power-law.
Having clarified the generic origin of power-law distribution for $c$, we
now turn our attention to the productivity dispersion across workers.
We must extend the theory of the stochastic macro-equilibrium explained in section II under the assumption that $\pfc$ is the power-distribution.  

In order to develop the new theoretical framework, it is 
better to adapt the continuous notation 
we introduced  previously.
An example is

\begin{equation}
K\sum_{k=1}^\infty n(c_k) \longrightarrow K\int_0^\infty \pf dc.
\end{equation}
In the continuous model, equations (\ref{entropy}), (\ref{norm}) and (\ref{dn}) read
\begin{gather}
S=-K\int_0^\infty \left( p(c) \ln p(c) \right)\pf dc, \label{p0}\\
K\int_0^\infty p(c) \,\pf dc=1,\label{p1}\\
K\int_0^\infty c\, p(c) \,\pf dc=D,\label{p2}
\end{gather}
respectively, 
Here, we have replaced $\pk=n_k/N$ by continuous function $p(c)$.
Note that 
because the distribution of productivity \textit{across firms} is no longer 
uniform, but is $\pf$, the corresponding distribution \textit{across workers}, $\pe$ is 
\begin{equation}
\pe\equiv K p(c) \pf.
\end{equation}
Using $\pe$, we can rewrite equations (\ref{p0}), (\ref{p1}) and (\ref{p2})
as follows:
\begin{gather}
S=-\int_0^\infty \pe \ln\frac{\pe}{\pf}\, dc + [{\rm const.}] ,\label{pe0}\\
\int_0^\infty \pe \,dc=1,\label{pe1}\\
\int_0^\infty c \,\pe \,dc=D.\label{pe2}
\end{gather}
Here, $[{\rm const.}]$ in equation (48) is an irrelevant constant term.
 
We maximize $S$ (equation ~(\ref{pe0})) under two constraints
(\ref{pe1}) and (\ref{pe2}) 
by means of calculus of variation
with respect to $\pe$ to obtain
\begin{equation}
\pe = \frac1{Z(\beta)}\pf\, e^{-\beta c}.
\label{pee}
\end{equation}
Here, as in section II, $\beta$ is the Lagrangian multiplier for the aggregate demand constraint, 
~(\ref{p2}), and the partition function $Z(\beta)$ is given by 
\begin{equation}
Z(\beta)=\int_0^\infty \pf \,e^{-\beta c} dc.
\label{zbp}
\end{equation}
It is easy to see that constraint (\ref{pe1}) is satisfied.
Constraint (\ref{pe2}) now reads
\begin{equation}
D=\frac1{Z(\beta)}\int_0^\infty c \,\pf\, e^{-\beta c}dc
\label{dtilexp}.
\end{equation}
This equation is equivalent to:
\begin{equation}
D
=-\frac{d}{d\beta}\ln Z(\beta).
\label{dnbetacon}
\end{equation}
It is straightforward to see that
equations (\ref{pee}) and (\ref{zbp}) are the  
counter parts of equations (\ref{boltz}) and (\ref{parti}), respectively,
under the assumption that the productivity dispersion across firms is not uniform
but is $\pfc$.

Because the productivity distribution across firms, $\pf$ obeys the 
Pareto law, the relation between the aggregate demand $D$ and $\beta $
(or the temperature) is not so simple as shown in 
 equation (\ref{ayrel}),
but is, in general, quite complicated.
However, we can prove that the power exponent  $\beta$ is a
decreasing function of the aggregate demand, $D$.
Thus, the fundamental proposition that when the aggregate demand is high,
production factors are mobilized to firms and sectors with high productivity 
(Figure \ref{fig:ay0}),
holds true in the extended model as well.
We provide the proof in \ref{appapp}.

\figkokopdf{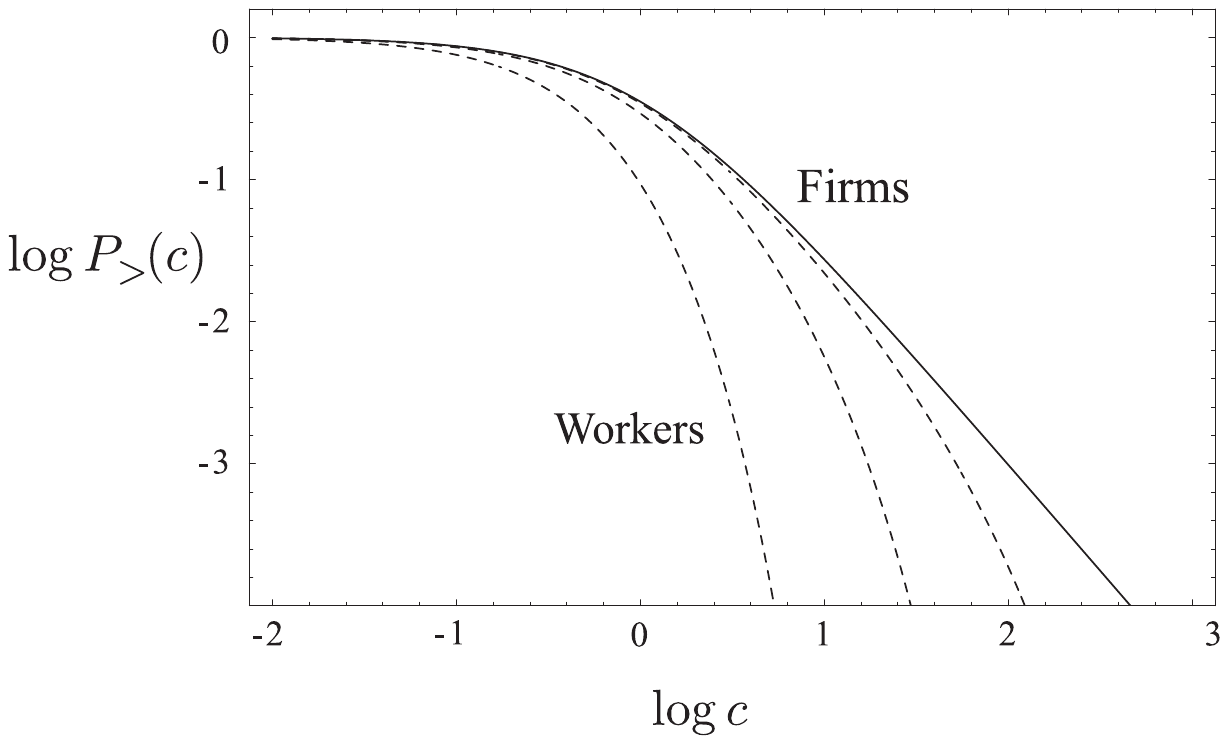}{\textsc{Productivity Distributions across Workers and Firms}
{\it Notes:} The solid curve is for firms with $\muf=1.5$, while the 
dashed curves are for workers with $\beta=0.01,0.1,1$, respectively. 
When $\beta$ is small enough,
the distribution across workers is close to that of
firms. As $\beta$ increases, the distribution across workers
is suppressed for large $c$ due to the exponential factor $e^{-\beta c}$.}{0.7}
{96 304 445 513}

Now, we explore productivity dispersion in this extended model.
The  productivity dispersion across workers, $\pe$ relates to that across firms, $\pf$
by way of equation (\ref{pee}).  
The latter obeys the power law.  
Figure \ref{fig:ay} shows 
examples of the productivity distributions across workers 
for several different values of $D$, given a $\pf$.
The solid curve is for firms with $\muf=1.5$.
The dashed lines are the corresponding distributions for
workers with different values of $D$.
All of them have strong suppression for large $c$ due
to the exponential factor $e^{-\beta c}$.
They are in stark contrast to the power-law.
Note that the power-law corresponds to a straight line in the figure
which shows the relation between $\ln P>(c)$ and $\ln c$.
The only way to reconcile the distribution (\ref{pee})
with the observed power-law is to assume an extremely small value of
$\beta$ so that the Boltzmann factor $e^{-\beta c}$ 
becomes close to one, and does not suppress $\pe$.
This trick, however, does not work because, in this case, it yields 
the Pareto index $\mue$ of the worker's productivity distribution
{\it equal to} that of the firm's distribution, $\muf$. 
It is inconsistent with the empirical observation that $\mue > \muf$ 
(Figure \ref{fig:ParetoIndex}).
Comparing Figures \ref{fig:employee2005_Nikkei} and \ref{fig:ay}, 
we  conclude that the extended model still fails to explain the observations.
We must seek a new theoretical framework. 

\subsection{Worker's Productivity Dispersion under Fluctuating Aggregate Demand}

The theoretical framework explained so far implicitly
assumes that the aggregate demand, $D$ is constant.
Plainly, this is an oversimplification;
$D$ fluctuates. 

Macro-system under fluctuations of
external environment can be analyzed with the help of
\textit{superstatistics} or
``statistics of statistics" in statistical physics
\citep{bc}.\footnote{There are several model cases where superstatistics was
applied successfully.
Among them, 
the Brownian motion of a particle going thorough changing environment
provides a good analogy to our case \citep{beck2}.}
In this theory, the system goes through changing
external influences, but is in equilibrium 
at the limited scale in time and/or space, in 
which the temperature may be regarded as constant and
the Boltzmann distribution is achieved.
In other words, the system is only locally in equilibrium;
Globally seen, it is
out of equilibrium.
In order to analyze such system, superstatistics 
introduces averaging over the Boltzmann factors. 
Depending on the weight function
used for averaging, it can yield various distributions,
including the power-law \citep{touchette2005as}.

GDP of a particular year is certainly a scalar constant when the year is over.  
However, it actually fluctuates daily if those fluctuations cannot be practically measured. 
Accordingly, macro environment surrounding firms keep changing almost continuously.  
Differences by region, industry and sector can also be taken care of by the super statistics approach.  
When the aggregate demand $D$ 
changes, the new 
stochastic process conditioned by new $D$ allows production factors to move to a different 
equilibrium. Averaged over various possible equilibria,
each of which depends on a particular value of $D$, 
the resulting distribution becomes different from any of each
equilibrium distribution.

Specifically, in superstatistics, the familiar Boltzmann factor, $\exp (-\beta  c)$ is replaced by
the following  weighted average:
\begin{equation}
B(c)=\int_0^\infty f(\beta) \, e^{-\beta c} d\beta,
\label{bc}
\end{equation}
Here, the weight factor $f(\beta)$ represents the changing macroeconomic environment.
Note that because $\beta $ is a monotonically decreasing function of $D$, the weight factor 
$f(\beta )$ corresponds to changes in the aggregate demand.
With this weight factor, the probability distribution of worker's productivity,
(\ref{pee}) is now replaced by the following:
\begin{equation}
\pe = \frac1{Z_B} \pf B(c).
\label{peess}
\end{equation}
Here, the partition function, $Z_B$ is also redefined as 
\begin{equation}
Z_B=\int_0^\infty \pf B(c) dc.
\label{zbpss}
\end{equation}

We now examine whether $\pe$ in equation (\ref{peess}) obeys the
power-law for high productivity $c$.
The integration in equation (\ref{bc}) is dominated by the small $\beta$
(high demand) region for large  $c$.
We assume the following behavior of the pdf $f(\beta)$ 
for $\beta\rightarrow 0$,
\begin{equation}
f(\beta) \propto \beta^{-\gamma} \quad (\gamma<1),
\label{fbeta}
\end{equation}
where the constraint for the parameter $\gamma$ 
comes from the convergence of the integration in equation (\ref{bc}).
The proportional constant is irrelevant because $\pe$ is normalized by $Z_B$.
This leads to the following  $B(c)$ for large $c$:
\begin{equation}
B(c) \propto   \Gamma(1-\gamma)\, c^{\gamma-1}.
\label{bcgamma}
\end{equation}
Substituting (\ref{bcgamma}) into Equation (\ref{peess}), we find that 
the productivity distribution across workers obeys the power-law;
\begin{equation}
\pe \propto c^{-\mue-1},
\label{mue}
\end{equation}
with
\begin{equation}
\mue=\muf-\gamma+1.
\label{muef1}
\end{equation}
Because of the constraint $\gamma <1$, this leads to 
the inequality
\begin{equation}
\mue > \muf.
\end{equation}
This agrees with our empirical observation (Figure \ref{fig:ParetoIndex} or stylized Fact II in 
section III.C).

Because $\beta$ is related to $D$ by way of equation (\ref{dnbetacon}),
the pdf $f_{\beta }(\beta)$ of $\beta$ is related to 
pdf $f_{D}(D)$ of the (fluctuating) $D$ as follows:
\begin{equation}
f_\beta (\beta)d\beta=f_{D}(D)dD.
\end{equation}
As is noted several times, small $\beta$ corresponds to  high aggregate demand, $D$. 
In particular, the following relation holds\footnote{The proof is to be given on request.} for
$\beta\rightarrow 0$:
\begin{equation}
\ac - D \propto
\begin{cases}
\, \beta & \mbox{for } 2<\muf; \\[7pt]
\, \beta^{\muf-1} & \mbox{for } 1<\muf<2.\\
\end{cases}
\label{dtilbeta}
\end{equation}
This leads to,\footnote{At $\muf=2$, we need additional logarithmic 
factors for $f(\beta)$, but the power of $\beta$ is essentially 
the boundary case between the above two, $\gamma=\delta$.} 
\begin{equation}
f_{D}(D)\propto\left(\ac-D \right)^{-\delta},
\label{deltadef}
\end{equation}
with 
\begin{equation}
\gamma-1=
\begin{cases}
\delta-1 
&\mbox{for }2<\muf;\\
(\muf-1)(\delta-1)
&\mbox{for }1<\muf<2.
\end{cases}
\label{gammarho}
\end{equation}
Here, the parameter $\delta$ is constrained by
\begin{equation}
\delta<1
\end{equation}
from the normalizability of the distribution of $f_{D}(D)$,
which is consistent with the constraint $\gamma<1$ and equation (\ref{gammarho}).

Equation (\ref{deltadef}) means that changes in the aggregate demand, 
$D$ follows the power-law. \citet{gabaix} indeed demonstrates 
that idiosyncratic shocks to the top 100 firms explain 
a large fraction (one third) of aggregate volatility for the U.S. economy. 
This is a characteristic of power-law. 
Any case, $D$ is not constant, but rather fluctuates, now.
Accordingly, the problem is not a relation 
between the productivity dispersion across workers and the level of $D$, 
but rather how the distribution depends on the way in which $D$ fluctuates.

\figkokopdf{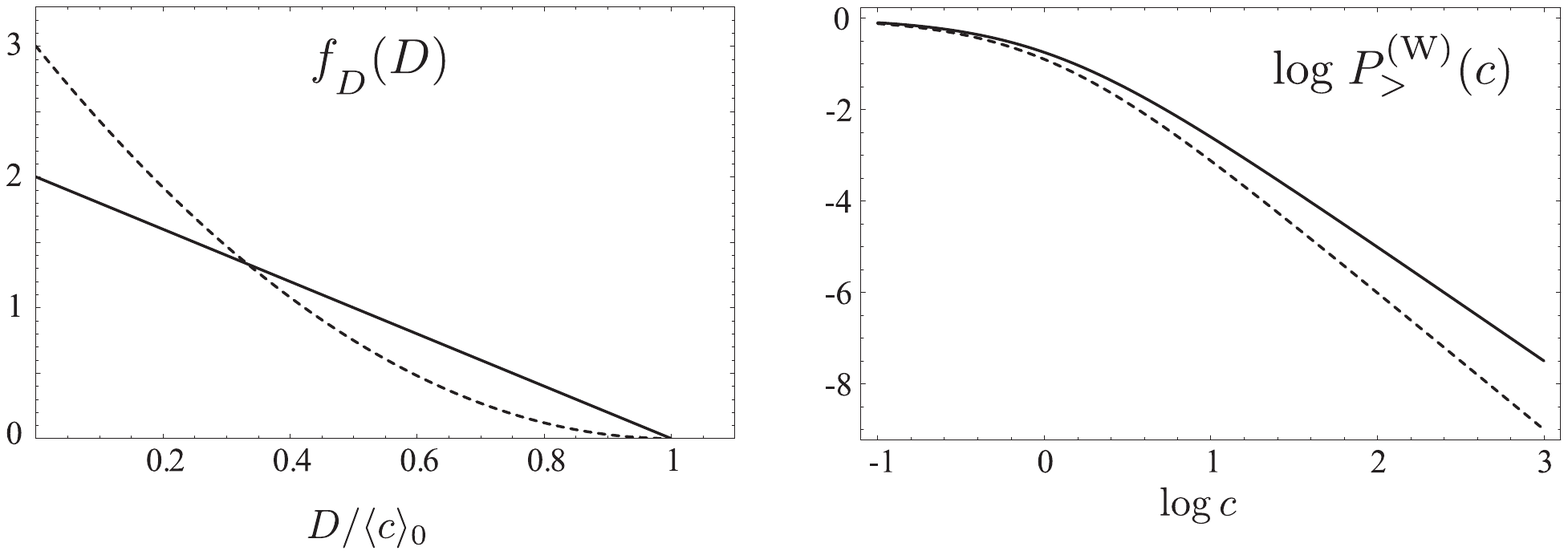}{\textsc{The Distribution of the Aggregate Demand $D$, 
$f_{D}(D)$ (left) and the Corresponding Cumulative Productivity Distribution $\pec$ (right)}
{\it Notes:} The solid curves are for $\delta=-1$ whereas the broken curves are for $\delta=-2$. 
The productivity distribution of firms is chosen to have $\muf=1.5$.}{1}
{144 187 702 382}

Figure \ref{fig:dpe} shows an example of the relation between 
the distribution of $D$, $f_D(D)$ near the upperbound $\ac$ 
and the cumulative productivity distribution of workers $\pec$. 
In the figure, the solid curves correspond to the large value of $\delta$ 
whereas the broken curves to the small value.
Figure \ref{fig:dpe} demonstrates  that
as the distribution of $D$ becomes skewed toward large $D$,
the tail of the productivity distribution becomes heavier. 
Roughly speaking, when $D$ is high, 
the productivity dispersion becomes skewed toward the higher level, 
and $vice$ $versa$.

Combining equations (\ref{muef1}) and (\ref{gammarho}),
we reach the following relation between the Pareto indices:
\begin{equation}
\mue= \begin{cases}
\muf-\delta+1 & \mbox{for } 2<\muf; \\[3pt]
(\muf-1)(-\delta+1)+\muf &\mbox{for } 1<\muf<2.
\end{cases}
\label{mumu}
\end{equation}
This relation between $\mue$ and $\muf$ 
is illustrated in Figure \ref{fig:mumu}.
As noted previously, because of the constraint $\delta<1$, 
equation (\ref{mumu}) necessarily makes $\mue$ larger than $\muf$.
This is in good agreement with our empirical finding. 
Incidentally, equation (\ref{mumu}) has a fixed point at $(\mue,\muf)=(1,1)$;
the line defined by equation (\ref{mumu}) always passes through this point irrespective of
the value of $\delta$. 
The Pareto index for firms is smaller
than that for workers, but it
 cannot be less than one,
because of the existence of the fixed point (1,1).

\figkokopdf{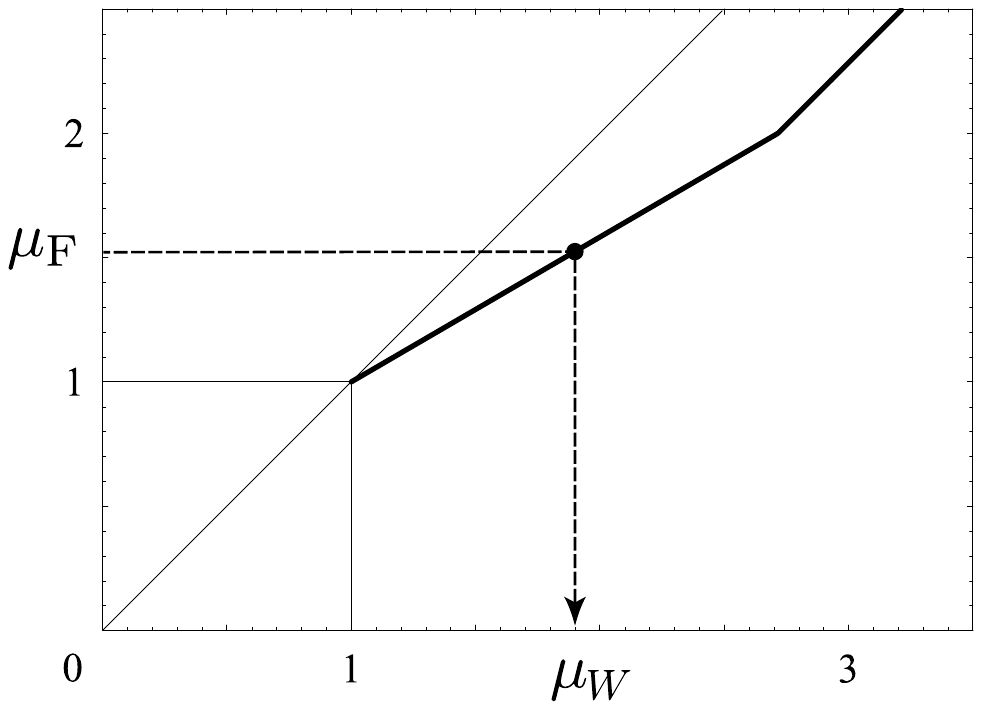}{\textsc{Relation between $\mue$ and $\muf$ (\ref{mumu})} 
{\it Notes:} The solid line is the relation (\ref{mumu}), and the filled circle 
is the data.}{0.5}{138 413 415 613}

The superstatistics framework presented above may apply
for any adjoining levels of aggregation;
Instead of applying it for workers and firms, we may 
apply it for firms and industrial sectors.
Then, we can draw the conclusion
 that as we go up from firms to industrial sectors, the
Pareto index again goes down, albeit for a  different value of 
$\delta$.
This is illustrated in Fig.\ref{fig:mumumu}.
Because of the existence of the point (1, 1), 
 as the aggregation level goes up,
the Pareto index is driven toward 1, but not beyond 1.
{\it At the highest
aggregation level, it is expected to be close to one}.
This is again in good agreement with our empirical finding that
the Pareto index of the industrial sector $\mus$ is 
 close to one (see Figure \ref{fig:ParetoIndex}).

\figkokopdf{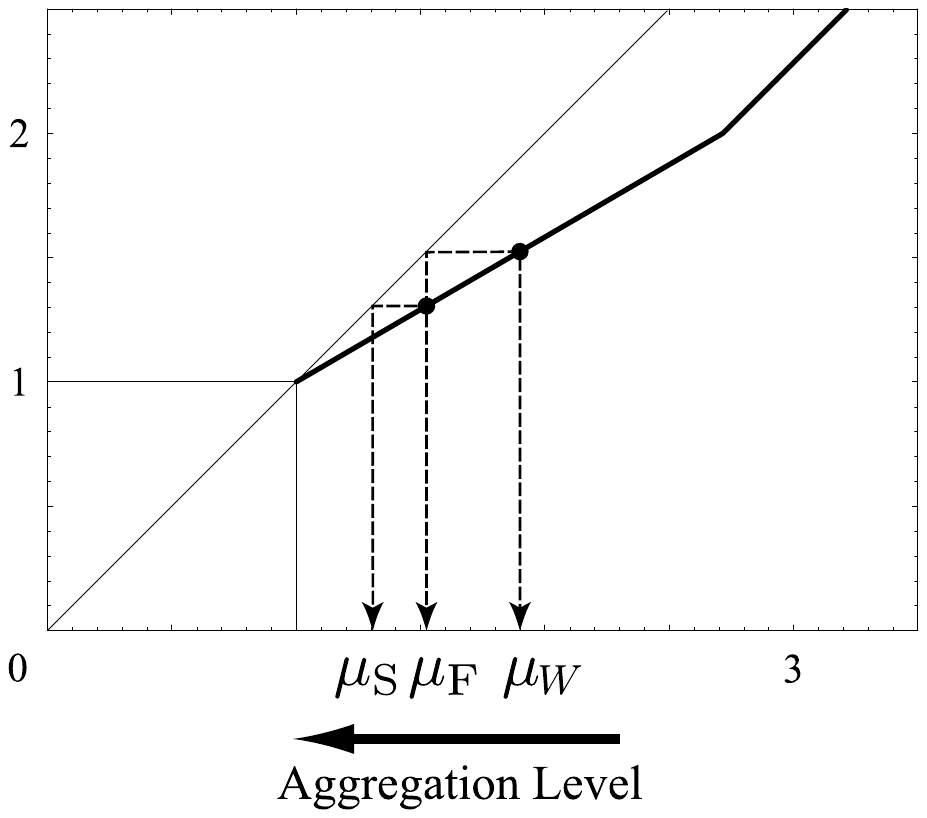}{\textsc{Changes of the Pareto Index
as the Aggregation Level changes in Two Steps, each with 
a Different Value of $\delta$}}{0.5}{153 382 415 613}

In summary, the superstatistics framework successfully explains two empirical findings
we have summarized in section III.C.
Furthermore, 
given the measured values of $\mue$ and $\muf$,
the relation (\ref{mumu}) can be used to determine the value of $\delta$:
\begin{equation}
\delta=
\begin{cases}
\muf-\mue+1 & \mbox{for } 2<\muf;\\[5pt]
\displaystyle \frac{\muf-\mue}{\muf-1}+1 &\mbox{for } 1<\muf<2.
\end{cases}
\label{deltais}
\end{equation}
The result is shown in Figure \ref{fig:delta}.
Recall that $\delta $ is the power exponent of the distribution of aggregate demand, $D$.  
Therefore, low $\delta$ means the relatively low level of the aggregate demand. 
In Figure \ref{fig:delta} we observe that 
the aggregate demand was high during the late 1980's, while beginning the 
early 90's, it declined to the bottom in 2000-2001, 
and then, afterward turned up.
It is broadly consistent with changes in the growth rate during the period. 

\figkokopdf{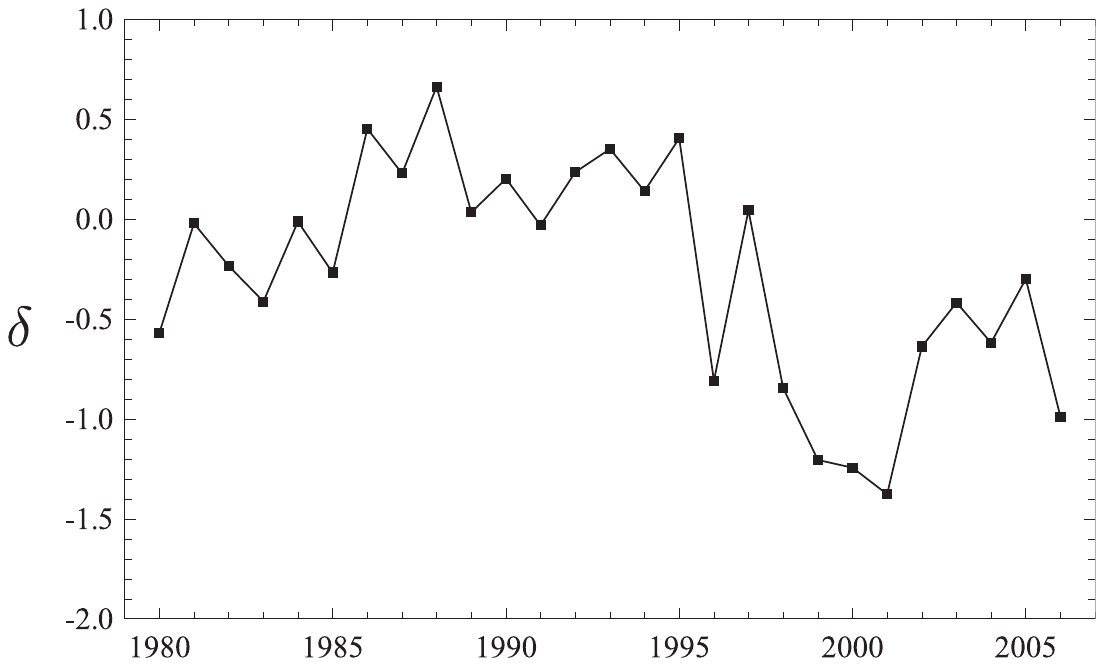}{\textsc{The values of $\delta$ calculated from Equation (\ref{mumu})
for the Japanese Listed Firms}}{0.7}{149 299 462 491}

\section{Implications}
The standard economic analysis takes it for granted that production factors 
move fast enough from low to high productivity firms and sectors, and that as a consequence, 
sooner or later they enjoy the same (highest) marginal productivity;  
Otherwise, it contradicts the concept of equilibrium.  
However, we have some evidences suggesting that there is always
productivity dispersion in the economy.  
As we referred to in Introduction, Mortensen (2003) analyzing wage dispersion argued that 
there is productivity dispersion across firms.  
\citet{okun2} 
also argued that a part of the reason why we obtain the Okun's law 
is ``the upgrading of workers into more productive jobs in a high-pressure economy.''  

In this paper, we have provided a solid foundation for Okun's argument.  
The most important point of Okun's argument and also our present analysis is that 
\textit{the allocation of production factors is not independent of the level of the aggregate demand}.  
Rather, it depends crucially on the aggregate demand.
As we explained in section II, the fundamental principle of statistical physics indeed 
tells us that it is impossible for production factors to achieve 
the same (highest) level of productivity.  
Rather, we must always observe the distribution of productivity in the economy.  
Moreover, the theory indicates, it is exponential distribution   (the Maxwell-Boltzmann),
the exponent of which depends inversely on the level of aggregate demand.  
As the aggregate demand rises, production factors are mobilized to high-productivity firms 
and sectors just as Okun argued.

In section III, we showed that there exist indeed the distributions of productivity
across workers, firms, and sectors.
A serious problem for the theory of stochastic macro-equilibrium is, however, that the observed distribution of 
productivity across workers is not exponential, but obeys the power-law.  
We reconciled this empirical observation with the basic theory by introducing 
 the assumption that productivity dispersion across firms obeys the power law 
rather than is uniform, 
and also another plausible assumption that the aggregate demand is not constant, 
but fluctuates.  

In section IV, first, we explained how the power distribution of productivity across firms arises in a simple stochastic model.  
To obtain the power law distribution, we need to make a crucial assumption that the higher the current level of productivity is, 
the greater the probability of either a unit increase or decrease of productivity is.  
Obviously, the probability of either \textit{a} birth or \textit{a} death is greater in a city with large population than in a small city.
By analogy, the above assumption seems quite natural.  

The ``size effect'' in productivity (or TFP) growth has been much discussed 
in endogenous growth theory 
because  its presence evidently contributes to endogenous growth.
See, for example, \citet{solow2000}.
In growth theory, an increase in productivity is mostly identified with pure technological progress so that 
it is directly linked to R \& D investment.  
However, to obtain the power distribution of productivity across firms, 
we must assume the significant probability of  \textit{decrease} in productivity.  
This suggests strongly that productivity changes facing firms  are caused not only by technical progress,
but also by \textit{the allocative disturbances to demand}. 
Incidentally, Davis, Haltiwanger, and Schuh (1996) report that unlike job creation, job destruction for an industry is not systematically related to total factor productivity (TFP) growth; 
Namely, job destruction occurs in high TFP growth industries as frequently as in low TFP growth industries (their Table 3.7 on page 52).
This fact also suggests the presence of the significant demand reallocation.

As Davis, Haltiwanger and Schuh (1996) rightly emphasize, 
the allocative demand shocks play a very important role in the macroeconomy.  
At the same time, the aggregate demand also plays a crucial role 
because the allocation of resources and production factors depends crucially 
on the level of aggregate demand. 
Put it simply, the frontier of production possibility set is a never-never land. 
The higher the level of aggregate demand is, the closer the economy is to the frontier.

\vspace{20pt}
\noindent
{\bf Acknowledgements}\\[3pt]
The authors would like to thank Dr.\ Y.\ Ikeda (Hitachi 
Research Institute) and Dr.\ W.\ Souma (NiCT and ATR)
for discussions.
Part of this research was supported by a grant from
Hitachi Research Institute. 
Yoshikawa's research is supported by RIETI. 
Yukawa Institute for Theoretical Physics at Kyoto University
supported this research in two ways; their computing facility
was used for part of the calculations and
discussions with participants at the YITP workshop YITP-W-07-16
{\it ``Econophysics III --Physical approach to social and economic phenomena"}
were helpful for completion of this work.

\appendix
\def\thesection{Appendix \Alph{section}}

\section{Notes on Data used in section III.A\label{appdata}}
As noted in section III, we used the Nikkei-NEEDS 
(Nikkei Economic Electronic Databank System) database 
for analysis of empirical distribution of the productivity.
This database is a commercial product available from \citet{nikkei}
and contains financial data of all the listed firms in Japan.
As such, it is a well-established and representative database, 
widely used for various purposes from research to practical business applications.
For our purpose, we used their 2007 CD-ROM version and extracted data for the period 
between 1980 and 2006. 
It covers some 1,700 to 3,000 firms and 4 to 6 million workers.

We have found that in certain cases, the productivity calculated is unrealistically large.
For example, firms that became stock-holding firms report huge reductions 
of the number of employees, while maintaining the same order of revenues in the year. 
This results in absurd values of labor productivity $c$ for that year.
Because of these abnormalities, we have excluded top-ten firms
in terms of the productivity each year.
This roughly corresponds to excluding firms with productivity $c > 10^9$ yen/person.
We experimented analyses with several different cuts, {\it i.e.},
with cutting top-twenty firms, and so on. 
The results obtained remained basically the same as reported in the main text.

The values of the Pareto indices $\mue, \muf, \mus$ given in this paper are determined by
fitting the data with the GB2-distribution 
described in \citet{actuarial} by the maximum likelihood method.

\section{The Relationship between Aggregate Demand $D$ 
and the exponent $\beta $\label{appapp}}
In this appendix, we prove the following
three basic properties (i)--(iii) of the 
temperature-aggregate demand relation (\ref{dnbetacon})
in Macro-Equilibrium:

\figkokopdf{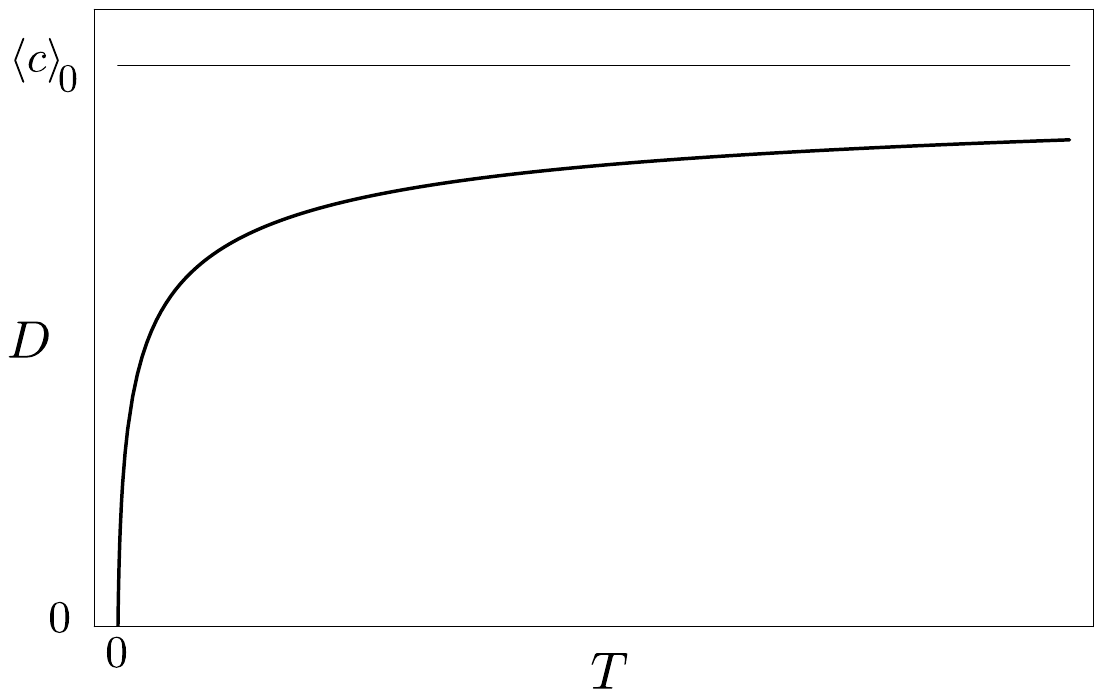}{\textsc{The Relation between the Aggregate Demand, $D$ 
and the Temperature, $T\ (=1/\beta)$}
{\it Notes:} For a productivity distribution of firms with $\muf=1.5$.}{0.6}
{296 237 609 433}

\begin{enumerate}
\item[(i)]
The temperature, $T=1/\beta$ is a monotonically increasing function of the aggregate demand, $D$.
We can prove it using equation (\ref{dnbetacon}) as follows:
\begin{align}
\frac{dD}{dT}=
-\frac1{T^2}\frac{dD}{d\beta}
=\beta^2\frac{d^2}{d\beta^2}\ln Z(\beta)
=\beta^2\left(\actn{2}-\act^2\right) \ge 0.
\end{align}
where $\actn{n}$ is the $n$-th moment of productivity defined as follows:
\begin{equation}
\actn{n}\equiv\frac1{Z(\beta)}\int_0^\infty c^n \pf \,e^{-\beta c}\,dc.
\label{momdef}
\end{equation}
Comparing (53) and (56), we know that $\act=D$.
This is a natural result. As the aggregate demand $D$ rises, 
workers move to firms with higher productivity.
It corresponds to the higher temperature
according to the weight factor $e^{-\beta c}$.
\item[(ii)]
For $T\rightarrow 0$ ($\beta\rightarrow\infty$),
\begin{equation}
D\rightarrow 0.
\end{equation}
This is evident from the fact that in the same limit the integration
in Eq.(\ref{dtilexp})
is dominated by $c\simeq 0$ due to the factor $e^{-\beta c}$, 
and the integrand has extra factor of $c$ compared to the
denominator $Z(\beta)$.
\item[(iii)]
For $T\rightarrow \infty$ ($\beta\rightarrow 0$),
\begin{equation}
D\rightarrow \int_0^\infty c\, \pf \,dc\ (=\ac). 
\end{equation}
This can be established based on the property (i) because 
$D=\act\rightarrow \ac$ as $\beta\rightarrow 0$ and 
$Z(0)=1$.
\end{enumerate}

An example of the relation between $D$ and $T\equiv 1/\beta$ is given in 
Figure \ref{fig:dnbeta}.

\ifx\undefined\bysame
\newcommand{\bysame}{\leavevmode\hbox to\leftmargin{\hrulefill\,\,}}
\fi

\end{document}